\documentclass[12pt]{iopart}
\usepackage{iopams}
\usepackage{graphicx}
\usepackage{cite}

\begin{document}
\title[Off-equatorial orbits in strong gravitational fields near compact objects -- II]
{Off-equatorial orbits in strong gravitational fields near compact objects -- II.\\[10pt]
\normalsize Halo motion around magnetic compact stars and magnetized black holes}

\author{J. Kov\'{a}\v{r}$^{1,2}$, O. Kop\'{a}\v{c}ek$^2$, V. Karas$^2$ and Z. Stuchl\'{\i}k$^1$}
\address{
$^1$Institute of Physics, Faculty of Philosophy and Science, Silesian University in Opava, Bezru\v{c}ovo n\'{a}m.~13, CZ-746\,01~Opava, Czech~Republic\\
$^2$Astronomical Institute, Academy of Sciences, Bo\v{c}n\'{i} II, CZ-141\,31~Prague, Czech~Republic}
\ead{Jiri.Kovar@fpf.slu.cz}

\begin{abstract}
Off-equatorial circular orbits with constant latitudes (halo orbits) of electrically charged particles exist near compact objects. In the previous paper, we discussed this kind of motion and demonstrated the existence of minima of the two-dimensional effective potential which correspond to the stable halo orbits.

Here, we relax previous assumptions of the pseudo-Newtonian approach
for the gravitational field of the central body and study 
properties of the halo orbits in detail. Within the general relativistic approach, we carry out our
calculations in two cases. Firstly, we examine the case of
a rotating magnetic compact star. Assuming that
the magnetic field axis and the rotation axis are aligned with each
other, we study the orientation of motion along the stable halo orbits. In the poloidal plane, we also discuss shapes of the related effective potential halo lobes where the general off-equatorial motion can be bound. Then we focus on the halo orbits
near a Kerr black hole immersed in an asymptotically uniform magnetic field of external origin.

We demonstrate that, in both the cases considered, the lobes exhibit two
different regimes, namely, one where completely disjoint lobes occur symmetrically 
above and below the equatorial plane, and another where the lobes are joined across the
plane. A possible application of the model concerns the structure of putative
circumpulsar discs consisting of dust particles. We suggest that the particles can acquire a small (but non-zero) net electric charge, and this drives them to form the halo lobes.
\end{abstract}


\maketitle

\section{\label{Sec:Intro}Introduction}
Motion of matter near compact stars and black holes has been discussed 
thoroughly in textbooks~\cite{Chandrasekhar,Mis-Tho-Whe:1973}.  In the
test-particle  approximation, the hydrodynamical terms are neglected, as
expected in circumstances when the medium is rarefied and the mean free
path is comparable with the typical length-scale of the system~\cite{Pra:1980,Bic-Stu-Bal:1989,Stu-Bic-Bal:1999}. The interplay between gravitational attraction and the electromagnetic 
attraction or repulsion is essential for characteristics of the 
motion, namely, its stability properties. Motivation for 
these studies arises from the problem of motion and acceleration of
matter (charged particles or dust grains)~\cite{Sen:1997,Vok-Kar:1991a,Fel-Sor:2003,Cal-Fel-Fab-Tur:1982,Stu-Hle:1998}.

Compact stars can be imagined as being endowed with magnetic dipoles
that are anchored in the stars and co-rotate with them~\cite{Mestel:1999,Sen:1997,Rez-Ahm-Mil:2001}. 
Rotation gives rise to the electric component of the field, and we
assume that  the resulting structure dominates over the small-scale
turbulent fields. On the other hand, uncharged black holes do not support their own intrinsic magnetic fields; they can only be embedded in external fields~\cite{Pet:1974,Wal:1974}. 
In general, the gravitational and electromagnetic fields interact with each other and the result is determined through the Einstein-Maxwell equations. To avoid such a complex description, one can also consider a test electromagnetic field influenced by the given gravitational field.

In both cases, magnetic compact stars as well as magnetized black holes\footnote{We use the term magnetic compact stars (magnetic stars) to specify neutron stars, Q-stars and hybrid stars as well, and the magnetized black holes for the black holes embedded in ordered magnetic fields.}, the interplay of strong gravity, rotation and magnetic field is capable of accelerating electrically  charged particles and, possibly, it can also introduce new families of  stable motion. The latter will be the main subject of the present paper. 

Recently, we have studied the existence of stable circular orbits with fixed latitudes of
charged particles moving off the equatorial plane (halo orbits)~\cite{Kov-Stu-Kar:2008,Stu-Kov-Kar:2009}. We assumed an axially
symmetric (aligned)  magnetic dipolar field and built our discussion on
classical studies~\cite{How-Hor-Ste:1999,Dull-Hor-How:2002} derived in
the context  of weak (Newtonian) gravitational fields, and designed for planetary studies. Particles on halo orbits are bound to the central object by a combined effect of gravitational and
electromagnetic  forces and they do not have to cross neither the
equatorial plane nor  the rotation axis. In the poloidal plane, the halo
orbits are located in a kind of lobes that are positioned symmetrically above and
below the equatorial  plane. The lobes may be disjoint or they may be
interconnected,  depending on the parameters. Our study~\cite{Kov-Stu-Kar:2008}, within  the pseudo-Newtonian approach~\cite{Stu-Kov:2008,Pac-Wii:1980,Abr:2009}, has
shown that the stable halo orbits  can indeed emerge also near
magnetic compact stars, where their structure is affected by strong
gravitational field. We have also examined the case of  charged and rotating
(Kerr-Newman) black holes and naked singularities.  Surprisingly, the
stable halo orbits do not exist above the horizon of Kerr-Newman black
holes. This suggests that the structure of the Kerr-Newman spacetime is in
several aspects quite special one, while the generic case of dipole-type
field usually allows for the motion along stable halo orbits.

Here, we extend our investigation of the halo orbits to the case of magnetic 
compact stars and magnetized black holes in general relativity.  To this end, we study the existence and orientation of stable halo orbits which surround the
minima of two-dimensional effective potential in the poloidal
($r,\theta$)-plane. We also deal with qualitatively different cases of the general, in the halo lobes bounded, off-equatorial motion (halo motion). Namely, we point out to the motion in two separated halo lobes extending symmetrically above and below the equatorial plane versus the motion in a single region arising from the merged lobes across the equatorial plane.
We also distinguish the cases when the halo lobes admit outflow of particles to the outer space or they allow the inflow onto the central body. 

In the space environment, dust particles can carry small electrostatic charges~\cite{Hor:1996}.
An astrophysical motivation to study the halo motion of such weakly charged
particles is the search for circumpulsar debris that could be formed from
a fall-back disc resulting from the fraction of the explosion ejecta
material that fails to escape~\cite{Cordes08,Muno06,Wang06}. The
anomalous X-ray pulsars~\cite{dunc92} represent a category of magnetars
around which the infrared observation indicate the presence of debris
dust discs, potentially relevant for our deliberation. These are young 
neutron stars in the category of magnetars with slow spin periods in a
range $\approx 2$--$12$ seconds. The minimum distance to which the dust can
reach is uncertain, however, part of it may enter the magnetosphere and
influence the current flows below the light cylinder, until the dust
evaporates. The potential relevance of the halo motion of dust particles
is based on the fact that these orbits occupy areas of the stable
motion. The irradiated dust particles acquire a small net electric
charge and modulate the source radiation at their characteristic
oscillation frequencies as they move in the halo lobes.

The paper is organized as follows. In section \ref{Sec:Formalism}, we
combine the standard formulation based  on the construction of the
super-Hamiltonian and the corresponding effective potential, with
perhaps a less frequent (in this context) formalism of forces in the
projected three-space~\cite{Abr-Nur-Wex:1995,Agu-etal:1996,Stu-Hle-Jur:2000,Kov-Stu:2007}. We maintain the
assumption about axial symmetry but we generalise ref.~\cite{Kov-Stu-Kar:2008} by considering the background of the magnetic compact 
star in a more consistent framework, i.e. in the Schwarzschild 
geometry with a test dipole-type rotating magnetic field~\cite{Sen:1997} (section~\ref{Sec:MagneticStar}). 
Furthermore, as a qualitatively different situation, we also consider a rotating black hole immersed in the uniform magnetic field~\cite{Wal:1974} (section {\ref{Sec:Wald}). We discuss
the astrophysical relevance in section~\ref{sec:Relevance} and then we conclude
the paper in section~\ref{sec:Conclusions}.
\section{\label{Sec:Formalism}Formalism}
According to the standard approach to a test particle
motion, we start by construction of the super-Hamiltonian~\cite{Mis-Tho-Whe:1973}
\begin{eqnarray}
\label{SuperHamiltonian}
\mathcal{H}=\textstyle{\frac{1}{2}}\;g^{ij}\,\Big(\pi_i-\tilde{q}A_i\Big)\,\Big(\pi_j-\tilde{q}A_j\Big),
\end{eqnarray}
where $m$ and $\tilde{q}$ are the rest mass and electric charge of the
particle, $\pi_i$ is the canonical momentum, and $A_i$ denotes the
vector potential related to the electromagnetic tensor by
$F_{ij}=A_{j,i}-A_{i,j}$. The particle motion is governed by Hamilton's
equations
\begin{equation}
\label{HamiltonsEquations}
\frac{{\rm d}x^i}{{\rm d}\lambda}=\frac{\partial \mathcal{H}}{\partial \pi_i},\quad \frac{d\pi_{i}}{d\lambda}=-\frac{\partial \mathcal{H}}{\partial x^{i}},
\end{equation}
where $\lambda=\tau/m$ is the affine parameter and $\tau$ is the proper
time.\footnote{We use the geometric system of units ($c=G=1$) and
a positive signature of the metric. In order to reduce the number of
parameters in our classification, we scale all quantities that have the
dimension of the power of length by the mass $M^*$ of the central
object. In this way we adopt the scaled dimensionless quantities. Thus,
our formulae become completely dimensionless. Furthermore, our 
attention is paid to the stationary and axially symmetric spacetimes, described by using the standard Boyer-Lindquist coordinates $x^i=(t,\phi,r,\theta)$, and endowed
with the magnetic fields which adopt the same symmetries.} The first Hamilton's equation of motion implies 
\begin{equation}
p^i\equiv\frac{{\rm d}x^i}{{\rm d}\lambda}=\pi^i-\tilde{q}A^i.
\end{equation}
The second Hamilton's equation ensures that the generalized momenta,
\begin{eqnarray}
\label{Momenta}
\pi_t&=&p_t+\tilde{q}A_t\equiv-\tilde{E},\quad
\pi_{\phi}=p_{\phi}+\tilde{q}A_{\phi}\equiv \tilde{L},
\end{eqnarray}
are constants of motion, reflecting the stationarity and axial
symmetry of the system. These are connected with
Killing vector fields $\eta^i=\delta^i_t$ and $\xi^i=\delta^i_{\phi}$.

We start by writing the normalization condition, $m^2=-g^{ij}p_i p_j$,
and defining the specific energy, angular momentum and charge, $E=\tilde{E}/m$,
$L=\tilde{L}/m$ and $q=\tilde{q}/m$, respectively. We find the
two-dimensional effective potential for the particle motion in the form~\cite{Mis-Tho-Whe:1973}
\begin{eqnarray}
\label{EffectivePotential_gen}
V_{\rm eff}=\frac{-\beta+(\beta^2-4\alpha\gamma)^{1/2}}{2\alpha},
\end{eqnarray}
where
\begin{eqnarray}
\label{EfectivePotential_parts}
\alpha=-g^{tt},\quad
\beta=2\Big[g^{t\phi}\Big(L-qA_{\phi}\Big)-g^{tt}qA_{t}\Big],\\
\gamma=-g^{\phi\phi}\Big(L-qA_{\phi}\Big)^2-g^{tt}q^2A_t^2+2g^{t\phi}qA_t\Big(L-qA_{\phi}\Big)-1,
\end{eqnarray}
reflecting the motion properties. 

Alternatively to the Hamilton's equations (\ref{HamiltonsEquations}),
one can describe the motion by the Lorentz equation
\begin{eqnarray}
\label{Lorentz}
u^k\nabla_k u^i=q\,F^{i}_{k}u^{k}.
\end{eqnarray}
For our purposes, we find this equation particularly well suited, when being rewritten in the formalism of forces~\cite{Abr-Nur-Wex:1995}. The forces formalism is based on the projection $h_{ik}=g_{ik}+n_i n_k$ of the Lorentz equation (\ref{Lorentz}) onto the three-dimensional hypersurface orthogonal to the four-velocity field of the Locally Non-Rotating Frames (LNRF)~\cite{bardeen},
\begin{eqnarray}
n^i=e^{-\Phi}(\eta^i+\Omega_{_{\rm LNRF}}\xi^i),\quad e^{2\Phi}=-(\eta^i+\Omega_{_{\rm LNRF}}\xi^i)(\eta^i+\Omega_{_{\rm LNRF}}\xi^i),
\end{eqnarray}
where the angular velocity $\Omega_{_{\rm
LNRF}}=-g_{t\phi}/g_{\phi\phi}$. In the case of a static spacetime,
$\Omega_{_{\rm LNRF}}=0$ and the LNRF become
static ones.

Let us consider an example of the halo motion -- the circular 
motion at constant latitude $\theta$ (outside
the equatorial plane). The four-velocity field of particles
uniformly  circling along the halo orbits can be decomposed as
\begin{eqnarray}
\label{A1}
u_{\rm h}^i=\gamma(n^i+v_{\rm h}\tau^i).
\end{eqnarray}
Here, $\gamma=(1-v_{\rm h}^2)^{-1/2}$ is the Lorentz factor,
$\tau^i=\xi^i\,g_{\phi\phi}^{-1/2}$ is the unit spacelike vector
orthogonal to $n^i$, along which the spatial velocity $v_{\rm
h}^i=v_{\rm h}\tau^i$ is aligned. Both vectors in the decomposition
(\ref{A1}) correspond to the base vectors of the standard orthonormal
tetrad attached to LNRF:  $n^i=e_{(t)}^i$ and
$\tau^i=e_{(\phi)}^i$. Thus, $v_{\rm h}$ is the orbital (azimuthal)
velocity measured with respect to LNRF.

Projection of the Lorentz equation, $h^k_j u^i\nabla_i u_k=
qh^i_j F_{ik}u^k$, can be written in the form
\begin{eqnarray}
\label{Force-equation}
\mathcal{G}_j+(\gamma v_{\rm h})^2\mathcal{Z}_j+\gamma^2v_{\rm h}\mathcal{C}_j=-q\gamma(\mathcal{E}_j+v_{\rm h}\mathcal{M}_j),
\end{eqnarray}
where the so-called mass and velocity independent parts of the
gravitational, centrifugal and Coriolis inertial forces, and the charge
and velocity independent parts of the electric and magnetic forces can be
expressed as 
\begin{eqnarray}
\mathcal{G}_j&=&-\partial_j\Phi,\\
\mathcal{Z}_j&=&\textstyle{\frac{1}{2}}g_{\phi\phi}^{-1}\,e^{-2\Phi}\Big(e^{2\Phi}\,\partial_{j}g_{\phi\phi}-g_{\phi\phi}\,\partial_j e^{2\Phi}\Big),\\
\mathcal{C}_j&=&g_{\phi\phi}^{-3/2}\,e^{-\Phi}\Big(g_{\phi\phi}\,\partial_{j}g_{t\phi}-g_{t\phi}\,\partial_jg_{\phi\phi}\Big),\\
\mathcal{E}_j&=&e^{-\Phi}\Big(\Omega_{_{\rm LNRF}}\partial_jA_{\phi}+\partial_j A_t\Big),\\
\mathcal{M}_j&=&g_{\phi\phi}^{-1/2}\,\partial_j A_{\phi},
\end{eqnarray}
where only the radial and latitudinal components are nonzero. 

Our investigation of the existence and orientation of the stable halo
motion is based on the following approach. First, we notice that 
loci of the halo orbits (stable as well as unstable) correspond to the
stationary points of the effective potential (\ref{EffectivePotential_gen}). We are interested in the stable motion mainly, i.e. in those orbits which satisfy the conditions for the local
minima of the potential,
\begin{eqnarray}
\label{Vcon1}
\partial^2_r V_{\rm eff}(r,\theta;p,L_{\rm h},q_{\rm h})>0,\\
\label{Vcon2}
\det{{\rm H}(r,\theta;p,L_{\rm h},q_{\rm h})}>0.
\end{eqnarray}
Here, H is the Hessian matrix, $p$ is a parameter characterizing the
compact object (angular velocity $\Omega$ in the case of magnetic star, spin parameter $a$ in the case of Kerr black hole),  and $L_{\rm h}$ and $q_{\rm h}$ are the specific angular
momentum and charge which are characteristic for a certain halo orbit. 

According to the forces formalism, we can determine $L_{\rm h}$
and $q_{\rm h}$ from the balance equations\footnote{Note that alternatively to equations (\ref{ForcesBalance})--(\ref{ForcesBalance2}), the specific angular momentum $L_{\rm h}$ and charge $q_{\rm h}$ can be determined also from the conditions $\partial_r V_{\rm eff}=0$ and $\partial_{\theta} V_{\rm eff}=0$. However, this routine is more complicated, especially in
our second case of the Kerr black hole in the uniform magnetic field.} 
\begin{eqnarray}
\label{ForcesBalance}
\mathcal{G}_r+(\gamma v_{\rm h})^2\mathcal{Z}_r+\gamma^2v_{\rm h}\mathcal{C}_r=-q_{\rm h}\gamma(\mathcal{E}_r+v_{\rm h}\mathcal{M}_r),\\
\label{ForcesBalance2}
\mathcal{G}_{\theta}+(\gamma v_{\rm h})^2\mathcal{Z}_{\theta}+\gamma^2v_{\rm h}\mathcal{C}_{\theta}=-q_{\rm h}\gamma(\mathcal{E}_{\theta}+v_{\rm h}\mathcal{M}_{\theta}).
\end{eqnarray}
Eliminating $q_{\rm h}$ and assuming $0<\theta<\pi$, $\theta\neq\pi/2$, we get the cubic equation 
\begin{eqnarray}
\label{Cubic}
Av_{\rm h}^3+Bv_{\rm h}^2+Cv_{\rm h}+D=0,
\end{eqnarray}
where
\begin{eqnarray}
A&=&\mathcal{M}_{\theta}(\mathcal{G}_r-\mathcal{Z}_r)+\mathcal{M}_{r}(\mathcal{Z}_{\theta}-\mathcal{G}_{\theta}),\\
B&=&\mathcal{E}_{r}(\mathcal{G}_{r}-\mathcal{Z}_{r})+\mathcal{E}_{r}(\mathcal{Z}_{\theta}-\mathcal{G}_{\theta})
+\mathcal{C}_{\theta}\mathcal{M}_{r}-\mathcal{C}_{r}\mathcal{M}_{\theta},\\
C&=&\mathcal{C}_{\theta}\mathcal{E}_{r}-\mathcal{C}_{r}\mathcal{E}_{\theta}+\mathcal{G}_{\theta}\mathcal{M}_{r}-\mathcal{G}_{r}\mathcal{M}_{\theta},\\
D&=&\mathcal{E}_{r}\mathcal{G}_{\theta}-\mathcal{E}_{\theta}\mathcal{G}_{r}.
\end{eqnarray}
The cubic equation (\ref{Cubic}) has in general three complex solutions,
\begin{eqnarray}
\label{Velocity}
v_{\rm h,i}=v_{\rm h,i}(r,\theta;p),
\end{eqnarray}
where ${\rm i}\in \{\rm {I,II,III}\}$. These can represent values of
orbital velocities of the charged particles moving along the halo orbits.
The corresponding specific charges can be derived from one of the
equations (\ref{ForcesBalance})--(\ref{ForcesBalance2}). We find 
\begin{eqnarray}
\label{Charge}
q_{\rm h,i}=\frac{\mathcal{G}_r(v_{\rm h,i}^2-1)-v_{\rm h,i}\,(\mathcal{C}_r+v_{\rm h,i}\,\mathcal{Z}_r)}{(\mathcal{E}_r+v_{\rm h,i}\,\mathcal{M}_r)\left(1-v_{\rm h,i}^2\right)^{1/2}}.
\end{eqnarray}
Finally, the three values of the specific angular momentum are
\begin{eqnarray}
\label{Angular}
L_{\rm h,i}=\gamma\,v_{\rm h,i}\,g_{\phi\phi}^{1/2}+q_{\rm h,i}\,A_{\phi}.   
\end{eqnarray}

Searching systematically through the parameter space
$(r\times\theta\times p)$, we can now investigate the validity of 
conditions (\ref{Vcon1}), (\ref{Vcon2}) and $v_{\rm h,i}\in R$, $|v_{\rm
h,i}|<1$, necessary and sufficient for the existence of the stable halo
orbits. 
\section{\label{Sec:MagneticStar}Magnetic star}
To a good approximation, the gravitational field outside a compact star is described by Schwarzschild metric~\cite{Mis-Tho-Whe:1973}
\begin{eqnarray}
\label{MetricSchw}
{\rm d}s^2&=&-\left(1-2r^{-1}\right){\rm d}t^2+\left(1-2r^{-1}\right)^{-1}{\rm d}r^2+r^2({\rm d}\theta^2+\sin^2{\theta}{\rm d}\phi^2).
\end{eqnarray}
Naturally, the gravitational field of rotating compact star differs from the Schwarzschild metric (in the case of slow rotation it is well determined by the Hartle-Thorne metric~\cite{hartle68}). However, one does not need to consider this discrepancy, taking advantage of the simple analytical form of the Schwarzschild metric element that captures the essential properties of the motion not only near non-rotating black holes, but 
above the surface of ultra-compact stars and farther away from slowly rotating black holes as well.
The test dipole magnetic field rotating with angular velocity $\Omega$ in the Schwarzschild geometry can be expressed in terms of the vector potential \cite{Sen:1997}
\begin{eqnarray}
\label{VectorPotentialMS}
A_{t}=\textstyle{\frac{3}{8}}\Omega\mathcal{M}\mathcal{R}\sin^2{\theta},\quad A_{\phi}=-\textstyle{\frac{3}{8}}\mathcal{M}\mathcal{R}\sin^2{\theta},
\end{eqnarray}
where 
\begin{equation}
\mathcal{R}=2+2r+r^2\ln{\left(1-2r^{-1}\right)}.
\end{equation}
The related dipole magnetic moment is~\cite{Bak-etal:2010}
\begin{eqnarray}
\label{Magnetic}
\mathcal{M}=\frac{4R^{3/2}\left(R-2\right)^{1/2}}{6(R-1)+3R\,(R-2)\,\ln{\left(1-2R^{-1}\right)}}\;B_{0}.
\end{eqnarray}

This configuration can be considered as a model of magnetic compact
star (radius $R$ and angular velocity $\Omega$) endowed with an 
aligned co-rotating magnetic field of strength $B_{0}$. The latter 
is measured with respect to the standard orthonormal tetrad of
static observers on the surface of the star in the equatorial plane. 
Rotation terms are neglected in the metric and the frozen-in condition
for the magnetic field is imposed, $F^i_j\, u^j_{\rm MF}=0$
(force-free approximation), 
where $u^i_{\rm MF}=(u^t_{\rm MF},u^{\phi}_{\rm MF},0,0)$, 
i.e. $u^{\phi}_{\rm MF}/u^t_{\rm MF}=\Omega$.

Employing equations (\ref{MetricSchw}) and (\ref{VectorPotentialMS}),
the effective potential (\ref{EffectivePotential_gen}) adopts the
form
\begin{eqnarray}
\label{EffectivePotential_MS}
V_{\rm eff}&=&-\textstyle{\frac{3}{8}}\,q\mathcal{M}\Omega\mathcal{R}\sin^2{\theta} \nonumber\\
&&+\left(1-2r^{-1}\right)^{1/2}\left[1+\left(\frac{L}{r\sin{\theta}}+\frac{3q\mathcal{M}\mathcal{R}\sin{\theta}}{8r}\right)^2\right]^{1/2}.
\end{eqnarray}
The only restriction following from the potential formula is $r>2$ (event horizon). However, the effective potential is not meaningful outside the region between the considered surface of the star and the light cylinder, arising due to the condition $g_{ik} u_{\rm MF}^i u_{\rm MF}^k<0$, thus being implicitly defined by the formula  
\begin{eqnarray}
\label{Plasma}
\Omega^2 r_{\rm lc}^3\sin^2{\theta_{\rm lc}}-r_{\rm lc}+2=0.
\end{eqnarray}

\begin{figure}[tbh!]
\centering
\includegraphics[scale=0.70, trim = 0mm 0mm 0mm 0mm, clip]{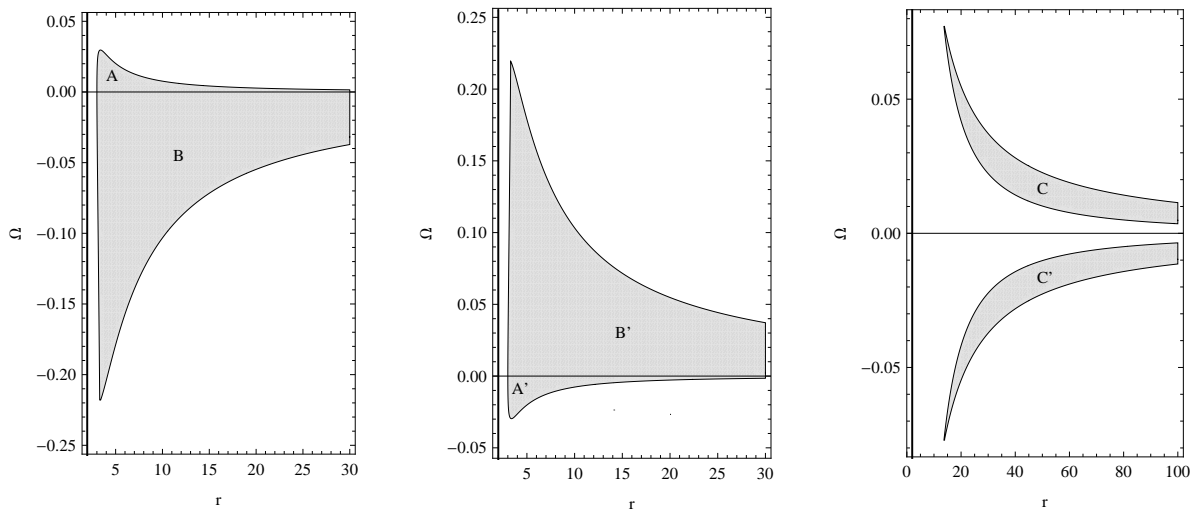}
\caption{The range and orientation of particle motion along stable halo
orbits  near a magnetic star. The star rotates with angular velocity
$\Omega$. This figure has been constructed for the latitude
$\theta=\pi/3$ and it represents a typical result for the effective
potential $V_{\rm eff}(r,\theta;L,Q,\Omega)$, based on the
behaviour of three orbital velocities $v_{\rm h,I}(r,\theta;\Omega)$, $v_{\rm
h,II}(r,\theta;\Omega)$ and $v_{\rm h,III}(r,\theta;\Omega)$ (roots of
equation (\ref{Cubic})).  These correspond to
the specific angular momenta $L_{\rm h,I}(r,\theta;\Omega)$,
$L_{\rm h,II}(r,\theta;\Omega)$, $L_{\rm h,III}(r,\theta;\Omega)$
and the effective specific charges $Q_{\rm h,I}(r,\theta;\Omega)$,
$Q_{\rm h,II}(r,\theta;\Omega)$, $Q_{\rm h,III}(r,\theta;\Omega)$. 
The gray areas correspond to the parameters for which the effective
potentials $V_{\rm eff}(r,\theta;L_{\rm h,I},Q_{\rm h,I},\Omega)$ (left
panel), $V_{\rm eff}(r,\theta;L_{\rm h,II},Q_{\rm h,II},\Omega)$
(middle panel)  and $V_{\rm eff}(r,\theta;L_{\rm h,III},Q_{\rm
h,III},\Omega)$ (right panel) develop local minima. Only the regions under the
light cylinder are considered. The thick vertical lines denote position of the event horizon at $r=2$. Description of the regions is summarized in table \ref{Tab:1}.} 
\label{Fig:1}
\end{figure}

\begin{table}[bth!]
\caption{Signs of quantities characterizing the motion along
stable halo orbits. Values of $r$ and $\Omega$ are taken from the
specific regions corresponding to figure \ref{Fig:1}.} 
{\small 
\begin{minipage}{.27\linewidth}
\begin{indented}
\item\begin{tabular}{@{}lcccc}
\br
Reg.&$v_{\rm h,I}$&$Q_{\rm h,I}$&$L_{\rm h,I}$\\
\hline
\hline
A&+&$-$&+\\
B&+&$-$&+\\
\br
\end{tabular}
\end{indented}
\end{minipage}
\begin{minipage}{.27\linewidth}
\begin{indented}
\item\begin{tabular}{@{}lcccc}
\br
Reg.&$v_{\rm h,II}$&$Q_{\rm h,II}$&$L_{\rm h,II}$\\
\hline
\hline
A$^\prime$&$-$&+&$-$\\
B$^\prime$&$-$&+&$-$\\
\br
\end{tabular}
\end{indented}
\end{minipage}
\begin{minipage}{.27\linewidth}
\begin{indented}
\item\begin{tabular}{@{}lcccc}
\br
Reg.&$v_{\rm h,III}$&$Q_{\rm h,III}$&$L_{\rm h,III}$\\
\hline
\hline
C&+&+&+\\
C$^\prime$&$-$&$-$&$-$\\
\br
\end{tabular}
\end{indented}
\end{minipage}
}
\label{Tab:1}
\end{table}

\subsection{Existence and orientation of halo orbits}
The general analysis in section \ref{Sec:Formalism} demonstrated that
the dipole magnetic field (\ref{VectorPotentialMS}) in the Schwarzschild
geometry (\ref{MetricSchw}) allows for the existence of the stable halo orbits. We took advantage of the fact that $q$ appears  in all
formulae as a product with $\mathcal{M}$, so we could introduce the
effective specific charge $Q=q\mathcal{M}$ instead of the
usual specific charge $q$. This decreases the number of
parameters. Only the parameter $p\equiv \Omega$ remains in the relations
for $v_{\rm h,i}$ (\ref{Velocity}), $L_{\rm h,i}$
(\ref{Angular}) and $Q_{\rm h,i}$, being expressed by the same formula as $q_{\rm h,i}$ (\ref{Charge}). 

The following conclusions about the existence of stable halo orbits arise
from figure \ref{Fig:1}. The stable halo orbits are possible for
particles which co-rotate with the dipole and whose charge satisfies
either the condition $Q_{\rm h}\Omega<0$ (regions A, B$^\prime$)
or $Q_{\rm h}\Omega>0$ (regions C, C$^\prime$). By co-rotation of
particles we mean $v_{\rm h}\Omega>0$, i.e. ${\rm sgn}(v_{\rm h})={\rm
sgn}(\Omega)$. Particles can also counter-rotate with the dipole along
stable halo orbits; in such case the charge must satisfy the condition
$Q_{\rm h}\Omega>0$ (regions A$^\prime$, B).  Thus, in the case of
the same orientation of the magnetic dipole and its rotation
($\mathcal{M}\Omega>0$), negatively charged particles moving along
stable halo orbits can only co-rotate, whereas positively charged
particles can either co-rotate or counter-rotate with the dipole. 

We point out that the above-mentioned results are general. Qualitatively
the same structure appears at any latitude, $0<\theta<\pi$,
$\theta \neq\pi/2$ (the assumption of $\theta=\pi/3$ in the figure
\ref{Fig:1} is not crucial). Note that $v_{\rm h}L_{\rm h}>0$
holds always for every stable halo orbit.

\subsection{Effective potential and classification of halo motion}
\label{effpot1}
As we can conclude from equation (\ref{EffectivePotential_MS}), the
effective potential is invariant under combinations of simultaneous sign
reversals maintaining
\begin{eqnarray}
{\rm sgn}(Q\,\Omega)={\rm const},\quad {\rm sgn}(\Omega L)={\rm const}.
\end{eqnarray}
At the event horizon, the potential diverges as 
\begin{eqnarray}
\label{Limit}
\lim_{r\rightarrow +2}V_{\rm eff}=\rm{sgn}(Q\,\Omega)\infty.
\end{eqnarray}
In the non-rotating case, the effective potential is invariant under the
signs reversals which satisfy
\begin{eqnarray}
{\rm sgn}(QL)={\rm const}.
\end{eqnarray}
 
The effective potential (\ref{EffectivePotential_MS}) exhibits a rich
spectrum of behaviour. Different topology of the effective potential determines different possible regimes of charged particles motion and yields the following classification. Note that locations of the surface of the star and of the light cylinder must be incorporated into the classification as well.  For instance, with respect to the different positions of
the light cylinder, we obtain an additional type IIIa$^\prime$ to the
type IIIa (although the behaviour of the potential itself is identical
in both the cases). Similarly, different positions of the surface of the
star bring the additional type Ia$^\prime$ to the type Ia, and the type
IIb$^\prime$ to the type IIb. On the other hand, the different types of the potential behaviour itself do not have to enrich the considered classification necessarily. For example, the different number of saddle points that occur below the star surface or beyond the light cylinder does not influence the topology between surface of the star and light cylinder, and is irrelevant for us. 

The family of different types of the effective potential can be parameterized by (conserved) energy of the moving particle. Starting at the energy corresponding to local
minima of the potential and rising gradually the energy level (and
keeping the other parameters constant), we first observe the gradually 
growing lobes of halo orbits, where particles are trapped. The following
classification emerges (see figures \ref{Fig:2}--\ref{Fig:3}).\footnote{Because of the complex spectrum of the different types, we present only the most representative cases in the following list.}

\begin{figure}[tbh!]
\centering
\includegraphics[scale=1, trim = 0mm 0mm 0mm 0mm, clip]{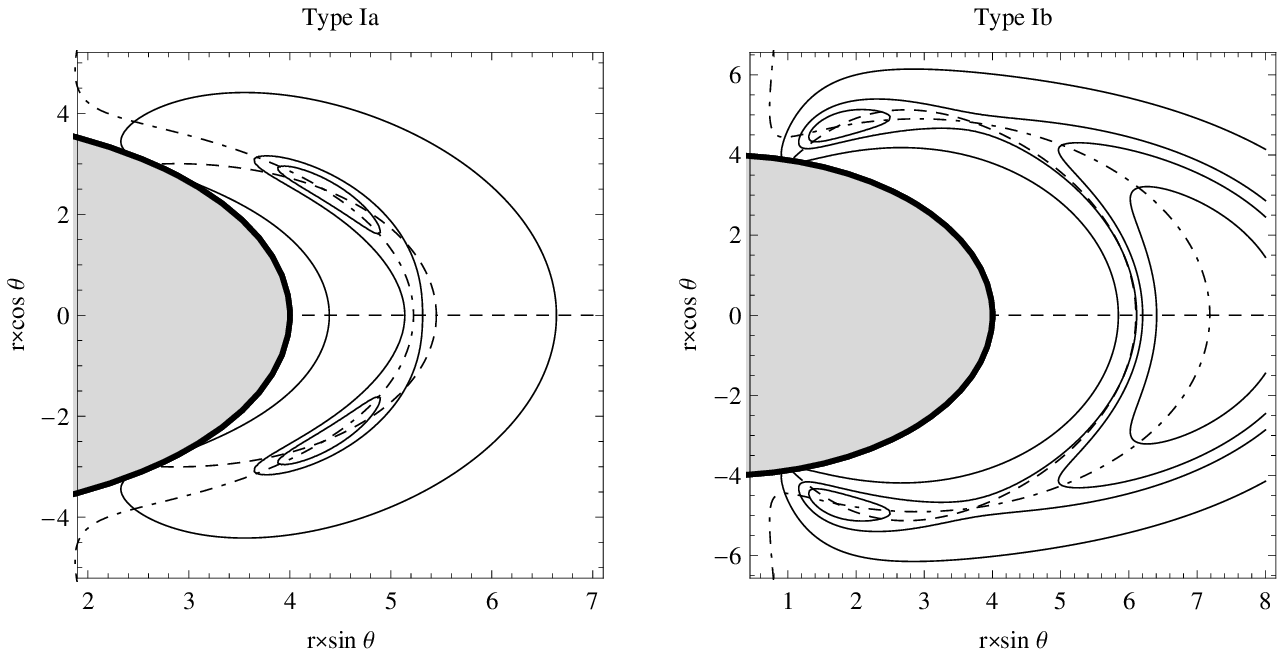}
\includegraphics[scale=1, trim = 0mm 0mm 0mm 0mm, clip]{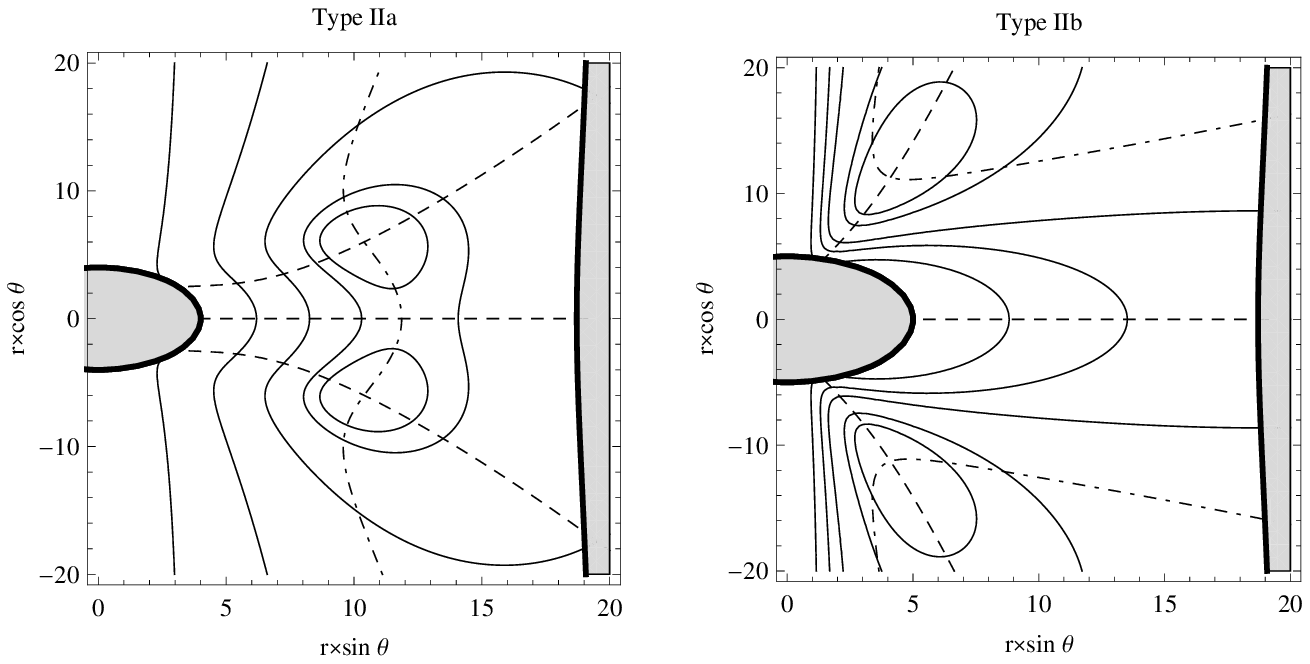}
\caption{Behaviour 
of effective potential $V_{\rm eff}$ for charged particles moving near
magnetic star. Qualitatively different types of behaviour of the
effective potential occur, with two local minima in the poloidal plane.
The lobes are limited by the radius of the star and the light cylinder
(thick curves limiting the gray areas). The equipotential surfaces are
shown by solid curves, whereas the dot-and-dashed and dashed lines
correspond to zeroes of $\partial_r V_{\rm eff}$ and $\partial_{\rm \theta}
V_{\rm eff}$, respectively. The presented types correspond to the
following combinations of parameters: {\bf Ia}: $Q\doteq -5.72$,
$L\doteq 0.876$, $\Omega\doteq 0.0115$, $R=4$ and $\theta_{\rm
h}=\pi/3, r_{\rm h}=5$ (parameters taken from region A in figure
\ref{Fig:1}); {\bf Ib}: $Q\doteq-14.6$, $L\doteq
0.343$, $\Omega=0.03$, $R=4$ and $\theta_{\rm h}=\pi/9, r_{\rm h}=5$; {\bf
IIa}: $Q\doteq 5.77$, $L\doteq -2.30$, $\Omega=0.0505$, $R=4$ and
$\theta_{\rm h}=\pi/3, r_{\rm h}=12$ (parameters taken from region
B$^\prime$ in figure \ref{Fig:1}); {\bf IIb}: $Q\doteq
25.2$, $L\doteq-0.788$, $\Omega=0.0505$, $R=5$ and $\theta_{\rm
h}=\pi/9, r_{\rm h}=12$.}
\label{Fig:2}
\addtocounter{figure}{-1}  
\end{figure}

\begin{figure}[t]
\centering
\includegraphics[scale=1, trim = 0mm 0mm 0mm 0mm, clip]{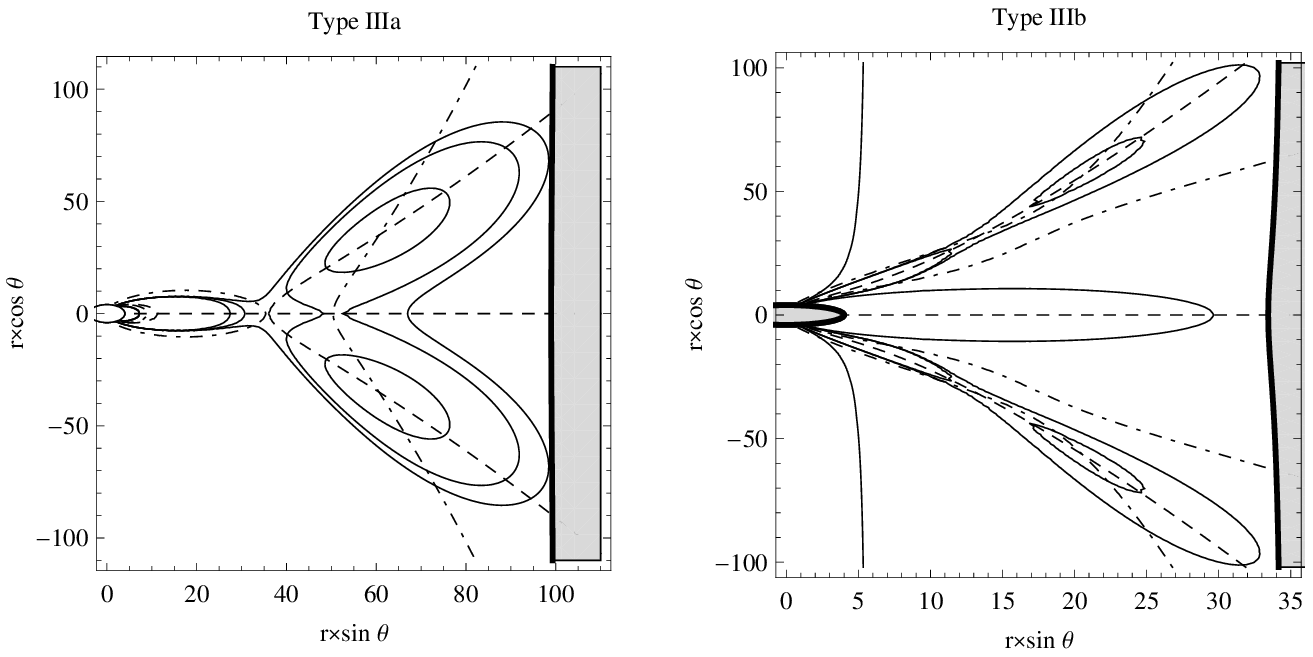}
\includegraphics[scale=1, trim = 0mm 0mm 0mm 0mm, clip]{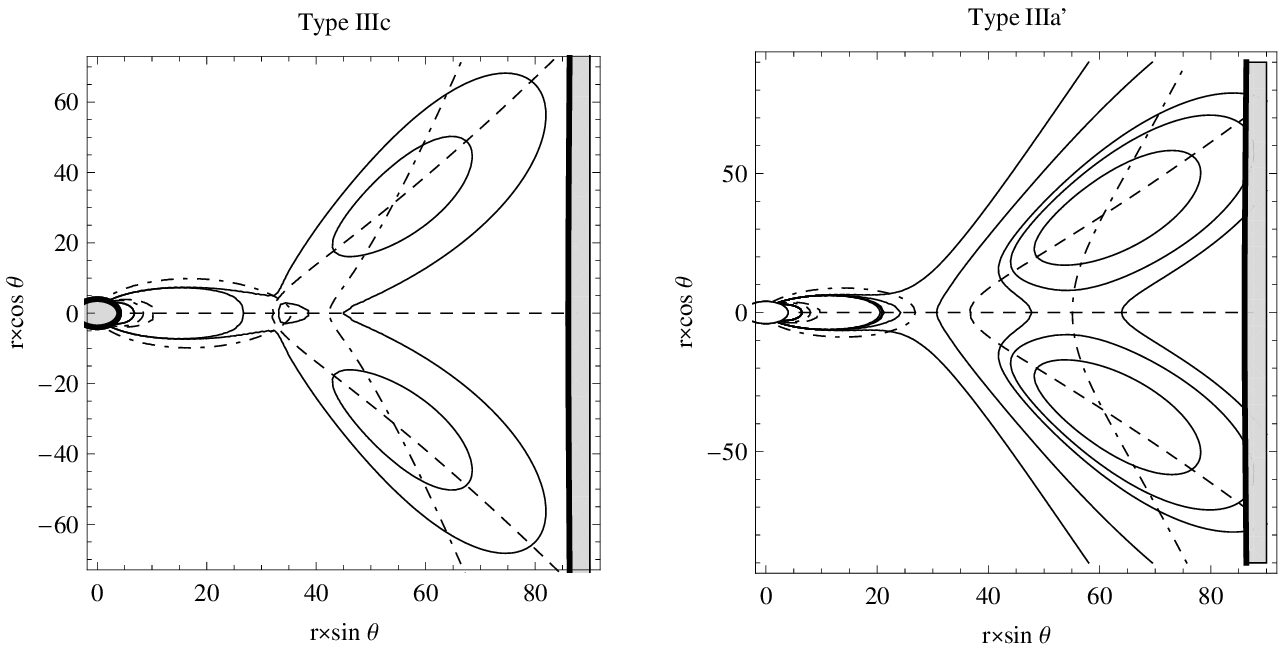}
\caption{(Continued.)
The presented types correspond to the following parameters: {\bf IIIa}:
$Q\doteq 53.4$, $L\doteq 6.58$, $\Omega=0.01$, $R=4$ and
$\theta_{\rm h}=\pi/3, r_{\rm h}=70$ (parameters taken from  region C in
figure \ref{Fig:1});  {\bf IIIb}: $Q\doteq 113$,
$L\doteq 2.43 $, $\Omega=0.029$, $R=4$ and $\theta_{\rm h}=\pi/9,
r_{\rm h}=60$;   {\bf IIIc}: $Q\doteq 45.9$, $L\doteq
6.25$, $\Omega\doteq 0.0115$, $R=4$ and $\theta_{\rm h}=\pi/3, r_{\rm h}=62.5$
(parameters taken from  region C in figure \ref{Fig:1}); {\bf
IIIa$^\prime$}: $Q\doteq 44.5$, $L\doteq 6.50$,
$\Omega=0.0115$ and $\theta_{\rm h}=\pi/3, r_{\rm h}=70$.}
\addtocounter{figure}{-1}
\end{figure}

\begin{figure}[t]
\centering
\includegraphics[scale=1, trim = 0mm 0mm 0mm 0mm, clip]{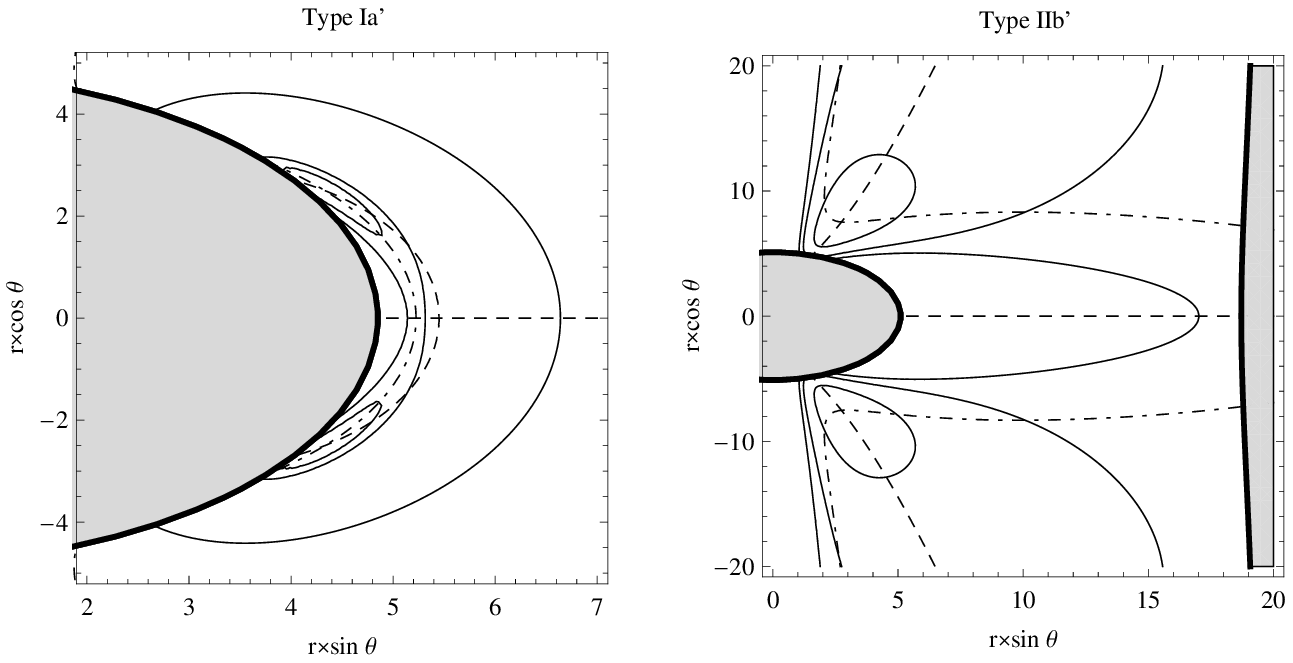}
\caption{(Continued.)
The presented types correspond to the following parameters: {\bf
Ia$^\prime$}: $Q\doteq -5.72$, $L\doteq 0.876$,
$\Omega\doteq 0.0115$, $R=4.85$ and $\theta_{\rm h}=\pi/3, r_{\rm h}=5$
(parameters taken from region A in figure \ref{Fig:1});  {\bf
IIb$^\prime$}: $Q\doteq17.1$, $L\doteq-0.617$,
$\Omega=0.0505$, $R=5.1$ and $\theta_{\rm h}=\pi/9, r_{\rm h}=8$.}  
\end{figure}

\subsubsection*{Type Ia.}
Halo lobes merge together once the energy level of the saddle point in
the equatorial plane is reached; particles moving within these merged
lobes can cross the equatorial plane. Later, when the energy is high
enough, particles start falling onto the surface of the magnetic star in
the regions where the merged potential lobes open, and the particles hit
the surface of the star. Increasing the energy further, the merged lobes
open out also on the outer side, crossing the light cylinder. This way
the particles can scatter away from the system.

\begin{figure}[tbh!]
\centering
\includegraphics[scale=1, trim = 0mm 0mm 0mm 0mm, clip]{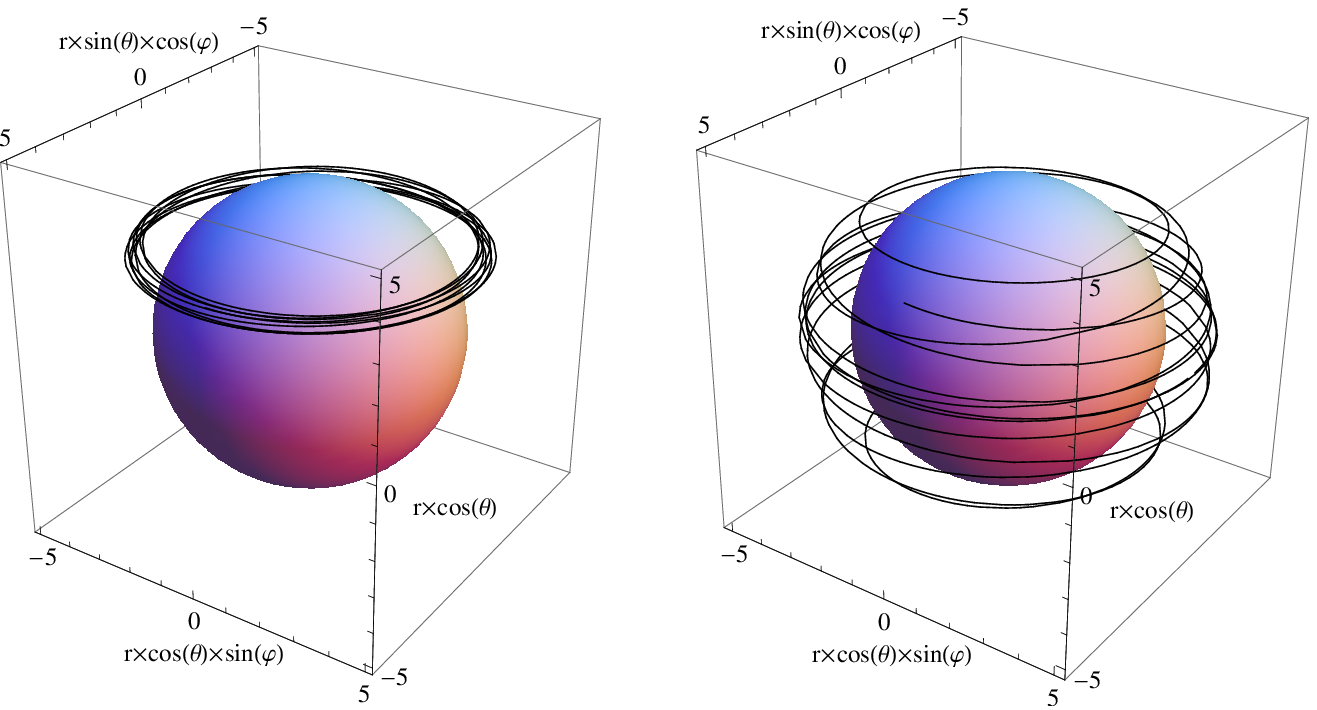}
\includegraphics[scale=1.1, trim = 0mm 0mm 0mm 0mm, clip]{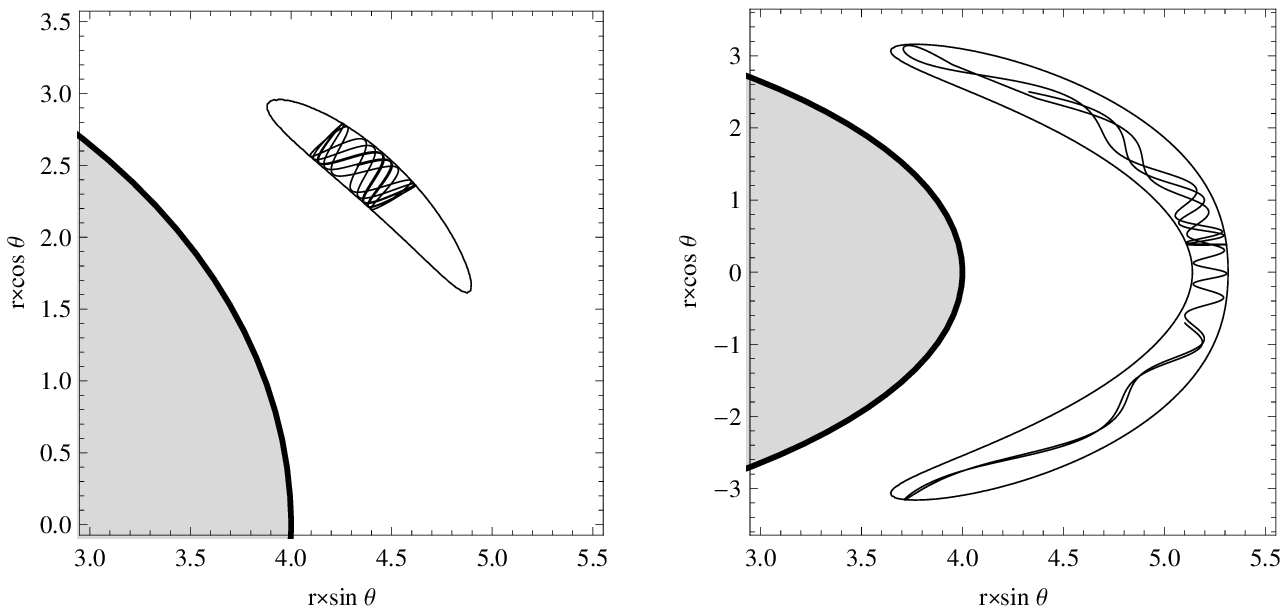}
\caption{Three-dimensional trajectories 
of charged particles (upper panels) with $Q\doteq -5.72$ and the
corresponding poloidal projection of the motion (lower panels), moving
in the separated and merged halo lobes in the vicinity of a magnetic
star with $R=4$ and $\Omega\doteq 0.0115$. The trajectories were
calculated numerically by using the Lorentz equation of motion
(\ref{Lorentz}) with the initial conditions $L\doteq 0.876$,
$\phi(0)=0$, $r(0)=5$, $\theta(0)=\pi/3$, and $E=0.8481$,
$u^{\theta}(0)=7\times10^{-4}$, $u^r(0)\doteq 0.021$ (motion in separated halo lobe), and
$E=0.8485$, $u^{\theta}(0)=7\times10^{-3}$, $u^r(0)\doteq 0.020$ (motion in merged
lobes).}
\label{Fig:3}
\end{figure}

\subsubsection*{Type Ib.} 
Halo lobes merge through saddle points out of the equatorial plane. 
When the energy is high enough, the merged lobes open out through the
surface of the star, and then also through the light cylinder. 
\subsubsection*{Type IIa.} 
Halo lobes merge through the saddle point in the equatorial plane. Then
again, the merged halo lobes open out through the light cylinder and,
finally, through the surface of the star. 
\subsubsection*{Type IIb.} 
Halo lobes open out through the light cylinder. When the energy is high
enough, the halo lobes merge through the saddle point located beyond the
light cylinder. Increasing the energy further, the opened and merged
lobes open out through the surface of the star. 
\subsubsection*{Type IIIa.} 
Halo lobes merge through the saddle point in the equatorial plane.
Later, when the energy is high enough, the merged lobes open out through
another saddle point in the equatorial plane. Finally, the lobes open
through the light cylinder.
\subsubsection*{Type IIIb.} 
Halo lobes open through saddle points out of the equatorial plane, and
later, when the energy is high enough, through the light cylinder as
well. The opened halo lobes merge through the saddle point in the
equatorial plane hidden beyond the light cylinder.
\subsubsection*{Type IIIc.} 
Halo lobes open out through saddle points out of the  equatorial plane
and merge this way with each other. Later, another corridor between the
lobes emerges through the saddle point in the equatorial plane. The
potential barrier between the saddle points can be overcome after
increasing the energy further. Finally, the merged and opened lobes open
out through the light cylinder.  
\subsubsection*{Type IIIa$^\prime$.}
Halo lobes open out through the light cylinder. Later the opened lobes
merge through the saddle point in the equatorial plane. Finally,
increasing the energy more, the opened and merged lobes open out through
the second saddle point in the equatorial plane. 
\subsubsection*{Type Ia$^\prime$.} 
Halo lobes open out through the surface of the star, then the halo lobes
merge through the saddle point in the equatorial plane, and, finally, by
increasing the energy the lobes open out through the light cylinder. 
\subsubsection*{Type IIb$^\prime$.} 
Halo lobes open out through the surface of the star, and later, when the
energy is high enough, through the light cylinder. Increasing the energy
further more, the opened lobes merge through the saddle point in the
equatorial plane hidden beyond the light cylinder.

\subsection{Features of the classification}
In the case of parallel orientation of the rotation and the magnetic moment of the dipole, the
behaviour of the potential of the types Ia, Ib and Ia$^\prime$ (serie I) determine
the motion of negatively charged particles which co-rotate along the
stable halo orbits. At the event horizon, the potential diverges as $V_{\rm eff} \rightarrow -\infty$, and there are two additional saddle points of the
potential near the event horizon (hidden under the surface of the star).
In the equatorial plane, the effective potential develops local maxima
(also hidden under the surface of the star) in addition to the saddle
point. Note that the extent of halo lobes can never be large (compared
to the radius of the black hole horizon) and they cannot be placed
farther off the star surface.

Behaviour of the types IIa, IIb and IIb$^\prime$ (serie II) determines the motion
of positively charged particles, counter-rotating along halo orbits. At the event horizon, the potential diverges as $V_{\rm eff} \rightarrow \infty$, and there are no other
stationary points hidden under the surface of the star, neither out of
the equatorial plane nor within the plane. A relatively high potential
barrier develops between the halo lobes and the star. Now, the halo lobes
can be of large extent and they can be placed farther away from the star
surface.

Finally, the behaviour of types IIIa, IIIb, IIIc and IIIa$^\prime$ (serie III)
determines the motion of positively charged particles, co-rotating along halo orbits. At the event horizon, the potential diverges as $V_{\rm eff} \rightarrow \infty$, and there
are no other stationary points out of the equatorial plane. However, in the equatorial plane, the
behaviour of the potential is more complicated.
In some cases, there can be even more than four stationary points.

There is an infinite potential barrier at the event horizon in the types of series II and III, and behaviour of the potential closer to the magnetic star can be complicated, especially in types of the serie III. But having the sufficiently high energy, particles can always fall onto the star surface, for the star surface always takes place above the event horizon.

The types Ia, IIa, Ia$^\prime$, IIIa and IIIa$^\prime$ are
characteristic for the `lower' stable halo orbits, taking place near the
equatorial plane. On the other hand, the types Ib, IIb and IIIb are
characteristic for the `upper' stable halo orbits which occur near the
rotation axis. Behaviour of the type IIIc represents an intermediate
situation.

\section{\label{Sec:Wald}Kerr black hole in uniform magnetic field}
In order to incorporate the large scale magnetic field in our
considerations, we employ Wald's test-field solution~\cite{Wal:1974}. 
The electromagnetic field is given in terms of the vector potential
\begin{eqnarray}
\label{waldpot1}
A_t=\textstyle{\frac{1}{2}}\,B_{0}\Big(g_{t\phi}+2a\,g_{tt}\Big)-\textstyle{\frac{1}{2}}\,\mathcal{Q}\,g_{tt}-\textstyle{\frac{1}{2}}\,\mathcal{Q},\\
\label{waldpot2}
A_{\phi}=\textstyle{\frac{1}{2}}\,B_{0}\Big(g_{\phi\phi}+{2a}\,g_{t\phi}\Big)-\textstyle{\frac{1}{2}}\,\mathcal{Q}\,g_{t\phi},
\end{eqnarray}
in the background of Kerr metric
\begin{eqnarray}
\label{KerrMetric}
{\rm d}s^2&=&-\frac{\Delta}{\Sigma}\Big({\rm d}t-a\sin{\theta}\,{\rm d}\phi\Big)^2+\frac{\sin^2{\theta}}{\Sigma}\Big[\left(r^2+a^2\right)\,{\rm d}\phi-a\,{\rm d}t\Big]^2\nonumber\\
&&+\frac{\Sigma}{\Delta}\,{\rm d}r^2+\Sigma\,{\rm d}\theta^2,
\end{eqnarray}  
where $\Delta=r^2-2r+a^2$, $\Sigma=r^2+a^2\sin^2{\theta}$ ($a$
is the rotational parameter). 
$\mathcal{Q}$ stands for the test charge of the black hole; the terms
containing $\mathcal{Q}$ may be identified with
the components of vector potential of the Kerr-Newman solution~\cite{Mis-Tho-Whe:1973}. Further, the asymptotic behaviour of components
(\ref{waldpot1})--(\ref{waldpot2}) justifies the identification of the
parameter $B_0$ with the strength of the uniform magnetic field into
which the Kerr black hole has been immersed. Wald~\cite{Wal:1974} has
shown that in the case of parallel orientation of the spin and the
magnetic field $B_{0}$, the black hole selectively accretes positive
charges (negative for the antiparallel orientation) until it is charged
to the equilibrium value
\begin{equation}
\label{waldcharge} \mathcal{Q}_{_{\rm W}}=2B_{0}\,a.
\end{equation}
We will adopt $\mathcal{Q}_{_{\rm W}}$ as a preferred value
of the charge in forthcoming examples.

\begin{figure}[tbh!]
\centering
\includegraphics[scale=0.8, trim = 0mm 0mm 0mm 0mm, clip]{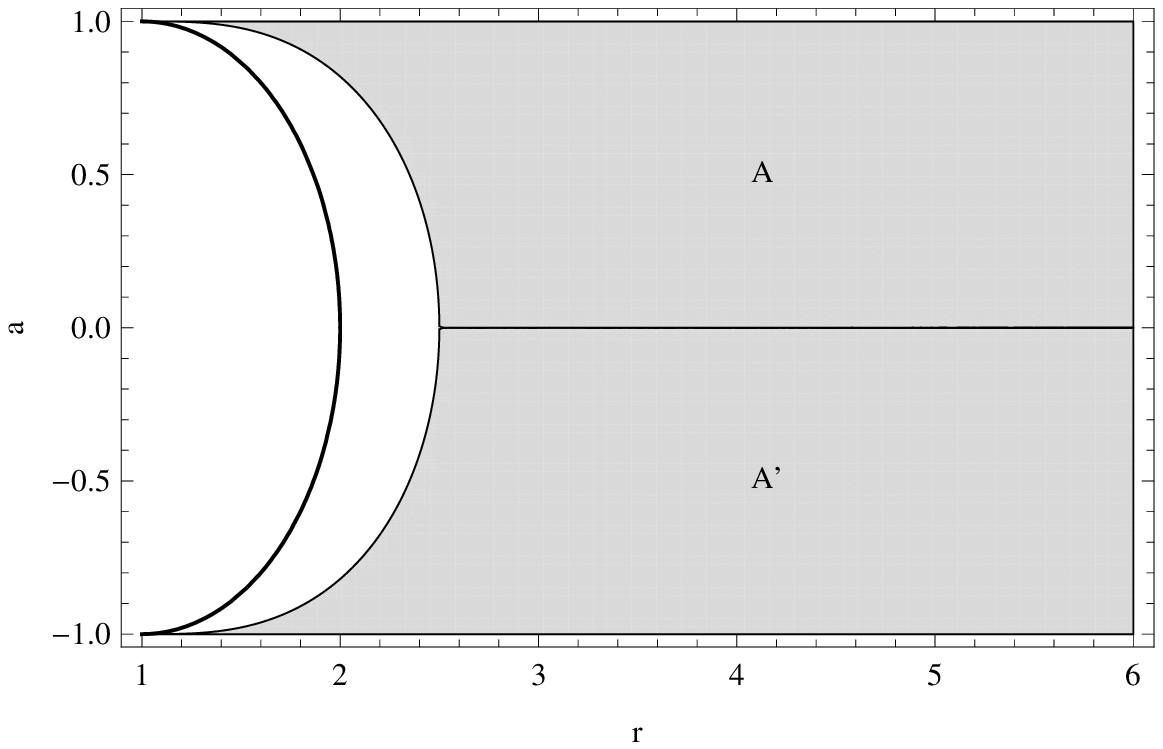}
\caption{The range and orientation of particle motion along stable halo
orbits near a Kerr black hole (with the spin $a$) embedded in a uniform  magnetic field. This figure has been constructed for the latitude $\theta=\pi/3$ and it represents a typical result for the effective potential $V_{\rm eff}(r,\theta;L,Q,a)$  based on the investigation
of the orbital velocity $v_{\rm h,I}(r,\theta;a)$ (root of equation (\ref{Cubic}) with $A=0$). The corresponding specific
angular momenta and charges are $L_{\rm h,I}(r,\theta;a)$ and $Q_{\rm h,I}(r,\theta;a)$. The gray regions correspond to the the
local minima of the effective potential $V_{\rm eff}(r,\theta;L_{\rm h,I},Q_{\rm h,I},a)$. The
thick solid curve shows the position of the event horizon at $r=1+(1-a^2)^{1/2}$. Further
classification is given in table~\ref{Tab:2}.}
\label{Fig:4}
\end{figure}
\begin{table}[tbh!]
\caption{Signs of quantities characterizing the motion along stable halo orbits. Values of $r$ and $a$ are taken from the specific regions corresponding to figure \ref{Fig:4}.}
{\small 
\begin{indented}
\item\begin{tabular}{@{}lcccc}
\br
Reg.&$v_{\rm h,I}$&$Q_{\rm h,I}$&$L_{\rm h,I}$\\
\hline
\hline
A&$-$&+&+\\
A$^\prime$&+&$-$&$-$\\
\br
\end{tabular}
\end{indented}}
\label{Tab:2}
\end{table}

\subsection{Existence and orientation of halo orbits}
Our analysis was described in general in section \ref{Sec:Formalism}. 
Its application to the case of Wald's test-field is shown in
figure~\ref{Fig:4}.  Again, we take the advantage of the fact that
$q$ appears only in the product with $B_0$, which leads us to
introduce the effective specific charge $Q=qB_{0}$. Then,
the only parameter $p\equiv a$ remains in the relations for $v_{\rm
h,i}$ (\ref{Velocity}), $L_{\rm h,i}$ (\ref{Angular}) and
$Q_{\rm h,i}$ being expressed by the same formula as
$q_{\rm h,i}$ (\ref{Charge}). Note that due to the use
of Wald's charge, the coefficient $A$ in equation (\ref{Cubic}) 
vanishes. Moreover, having now the quadratic equations, one of the two possible roots does not take the real values from the interval $(-1,1)$ in the considered black hole region above the outer event horizon. Thus, there is only one possible root $v_{\rm
h,I}(r,\theta;a)$ and the related
specific angular momentum $L_{\rm h,I}(r,\theta;a)$ and effective specific charge
$Q_{\rm h,I}(r,\theta;a)$ involved in the classification routine. 

In the region above the outer event horizon, the stable halo orbits are
possible only for particles which counter-rotate with respect to LNRF 
($v_{\rm h}a<0$) and charge of which satisfies the condition $Q_{\rm h}a>0$ (regions A, A$^\prime$).
The parallel orientation of the spin $a$ and the magnetic field $B_0$
($aB_0>0$) allows counter-rotation along stable halo orbits only for
positively charged particles. Moreover, in the potential minimum, there
is always $v_{\rm h}L_{\rm h}<0$.

\subsection{Effective potential and classification of halo motion}
We can conclude from formulae (\ref{KerrMetric}),
(\ref{EffectivePotential_gen}), (\ref{waldpot1}) and (\ref{waldpot2})
that the potential $V_{\rm eff}$ allows several combinations of
simultaneous sign reversals which follow from its symmetries. These have
to satisfy the conditions 
\begin{eqnarray}
{\rm sgn}(Qa)={\rm const},\quad {\rm sgn}(aL)={\rm const}.
\end{eqnarray}

The effective potential (\ref{EffectivePotential_gen}) does not reveal
as many possibilities of stable halo motion as we have seen in the
previous case of a magnetic star in section~\ref{effpot1}. Moreover, here, only the topology of the potential is crucial for the classification of the halo motion.  
Starting at energy of the potential local minima and rising its level
(keeping other parameters constant), we still observe the growing
halo lobes in which particles are trapped. The system can
be classified in the following way (see figure \ref{Fig:5}).

\begin{figure}[tbh!]
\centering
\includegraphics[scale=1, trim = 0mm 0mm 0mm 0mm, clip]{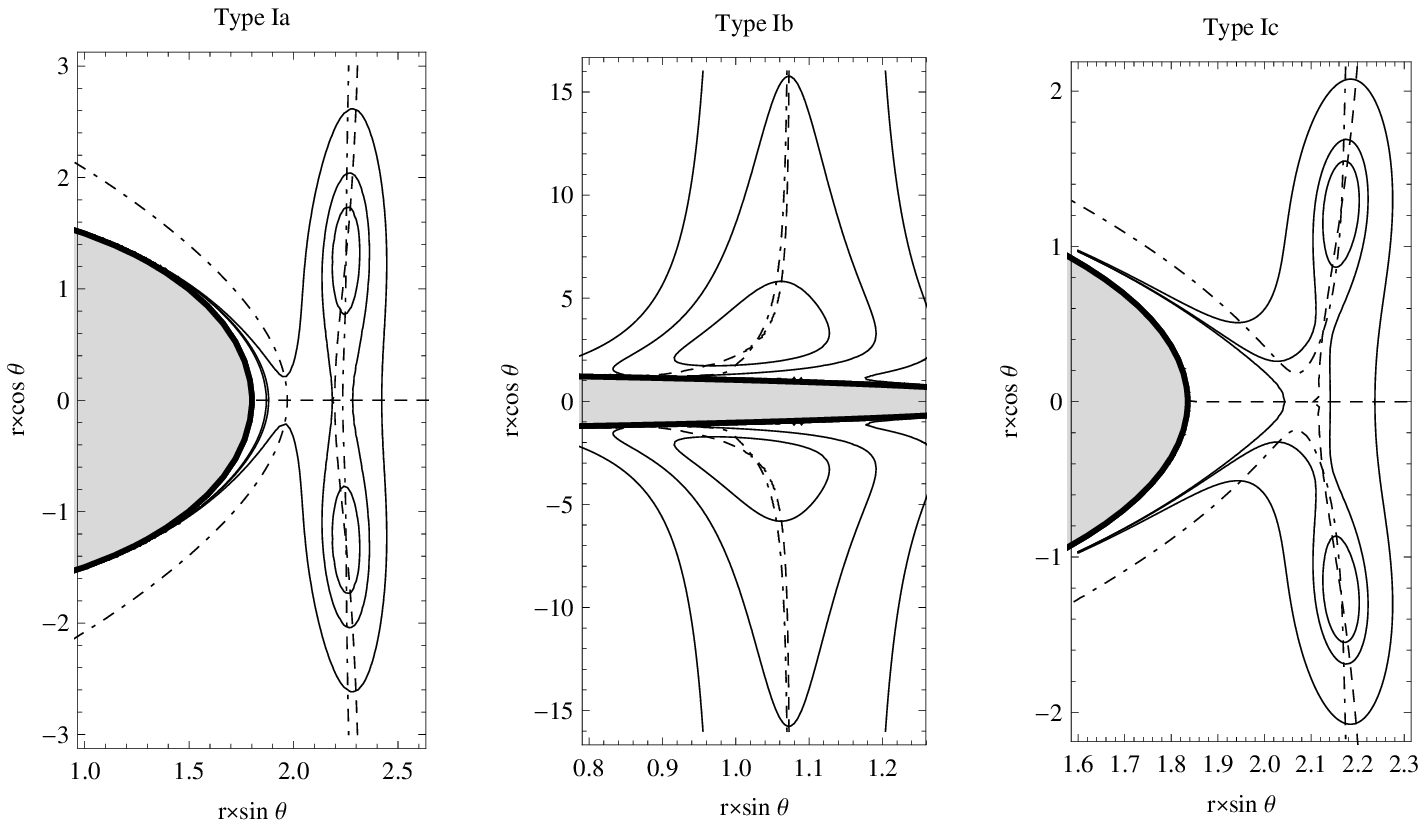}
\caption{Behaviour 
of the effective potential $V_{\rm eff}$ for charged particles moving
near Kerr black hole in the uniform magnetic
field. These are typical examples of behaviour of the potential with two
potential minima, again introduced in terms of the equipotential
surfaces. The dashed and dot-and-dashed curves correspond to zeros of
$\partial_{\theta} V_{\rm eff}$ and  $\partial_r V_{\rm eff}$,
respectively. The gray region corresponds to the interior of black hole,
limited by the outer event horizon.  The presented types correspond to
the following parameters: {\bf Ia}: $Q\doteq1.37$,
$L\doteq3.76$, $a=0.6$ and $\theta_{\rm h}=\pi/3, r_{\rm
h}=2.6$; {\bf Ib}: $Q\doteq 4.79$, $L\doteq2.76$, $a=0.9$
and $\theta_{\rm h}=\pi/9, r_{\rm h}=3$; {\bf Ic}: $Q\doteq 1.60$,
$L\doteq 4.02$, $a=0.55$ and $\theta_{\rm h}=\pi/3, r_{\rm
h}=2.5$. The parameters of all the presented types are taken from the
region A in figure \ref{Fig:4}.}
\label{Fig:5}
\end{figure}

\subsubsection*{Type Ia.} 
Halo lobes merge with each other when the energy level is reached of the
saddle point in the equatorial plane. Particles moving in these merged
lobes can cross the equatorial plane. When the energy is increased high
enough, particles can fall into the black hole, passing the next saddle
point in the equatorial plane, through which the merged halo lobes open
out. 
\subsubsection*{Type Ib.}
Halo lobes open out through the saddle points out of the equatorial
plane and the trapped particles can fall into the black hole.
\subsubsection*{Type Ic.} 
Halo lobes merge through two saddle points out of the equatorial plane
and they open out this way. Trapped particles can cross the equatorial
plane or fall into the black hole.
\\

In the case of the parallel orientation of the spin and the magnetic
field, the behaviour of all the presented types determines the motion of
positively charged particles counter-rotating along the stable halo
orbit. The behaviour of type Ia is characteristic for potentials
determining the `lower' stable halo orbits. Type Ib is characteristic
for the potential determining the `upper' stable halo orbits. Type Ic
exhibits an intermediate type.

In all the presented results here, we assume $\mathcal{Q}=2B_0a$. But we checked that the stable halo orbits are also possible in the uncharged case, $\mathcal{Q}=0$, finding no substantial change in the topology of effective potential
compared to the charged case.

\section{\label{sec:Relevance}Discussion: The astrophysical relevance of halo orbits}
In this paper, we have concentrated ourselves on aspects of the 
off-equatorial motion near magnetic compact stars and magnetized black holes.  Namely, we
discussed the existence of halo orbits, classified  them into several
categories and investigated their dependence  on the parameters. Despite
the obvious limitations of our  approach, one can recognize several
astrophysically relevant  features of the halo orbits.

We can list the main limitations of the present approach: (i)~Test-particle approximation was employed. This assumes that the  medium is
highly diluted and the mean free path is comparable  with the
characteristic size of the system (gravitational radius of the black
hole). (ii)~Schwarzschild metric was adopted for the external
gravitational field of the magnetic compact star and Kerr metric for the
rotating black hole, ignoring all other terms that could contribute  to
the gravitational field. Such contributions could in principle 
originate from the internal structure of the star and its  rotation, or
they could be caused by accreted matter outside  the central body.
(iii)~The rotating dipole or the asymptotically  uniform structure has
been imposed as two examples of the magnetic field and the associated
electric field. It was also assumed that the electromagnetic field does
not affect the spacetime metric, so it could be treated in the
test-field approximation.

The assumptions (ii) and (iii) are less critical because they are well
fulfilled under the typical astrophysical conditions. We considered two
different examples of the magnetic and gravitational fields, we found
that the halo orbits are possible in both.  The halo orbits are
determined by their location in the poloidal ($r$,$\theta$) plane, and
by values of other parameters, such as the magnetic field strength and
the specific charge of the particles. The assumption (i) is the crucial
one and it imposes the main restriction on the choice of the system
which can be described.

As for the numerical values of the mass-scaled (dimensionless) parameters $\Omega$, $a$ and
$Q$, we can check the adequate choice by looking at the domain of
the non-scaled quantities $\Omega^*=\Omega/M^*$, $a^*=aM^*$,
$B_0^*=B_0/M^*$, $q^*=q$, $R^*=RM^*$ and $M^*$,
characterizing real objects.

\subsection{Magnetic star}
In the following analysis, we assume masses and radii of magnetic compact stars in the ranges \mbox{$1\,\rm{M_{\odot}}\leq$$M^*$$\leq3\,\rm{M_{\odot}}$} and \mbox{$3$$M^*$$\leq R^* \leq 10$$M^*$}~\cite{Lat-Pra:2004}, and magnetic fields and rotation frequencies reaching $B_0^*=10^{10}\,\rm{T}$ and $f^*=10^3\,{\rm Hz}$ ($\Omega^*\doteq 6283\,{\rm rad/s}$)~\cite{Lat-Pra:2007,Bej-eta:2008,Hae-eta:2009}.\footnote{The precise determination of the mentioned limits is still of high interest, being very complex, dependent on the considered equation of state, etc. Moreover, not all the theoretical calculated models have their observable counterparts. For instance, as for the masses of neutron stars, the narrower range \mbox{$1.4\,\rm{M_{\odot}}\lesssim$$M^{*}$$\lesssim 2.5\,\rm{M_{\odot}}$} is widely stated for the astrophysically relevant neutron stars. As for the radii, most of the realistic equations of state imply the lower limit only $R^*\approx 3.5M^*$~\cite{Gle:1997}. On the other hand, the existence of extremely compact stars with $R^*<3M^*$ is also discussed. Some  models for the so-called Q-stars allow the lower limit even $R^*\approx 2.8M^*$~\cite{Bah-etal:1989,Mil-etal:1998,Stu-etal:2009}. In this paper, the limits are considered only for order estimates of the used dimensionless parameters, thus their exact values are of less importance here.} 

The limits of the used scaled parameters can be determined from the figure \ref{Fig:6}, which can be read in the following way.    
\begin{figure}[t]
\centering
\includegraphics[scale=1.2, trim = 0mm 0mm 0mm 0mm, clip]{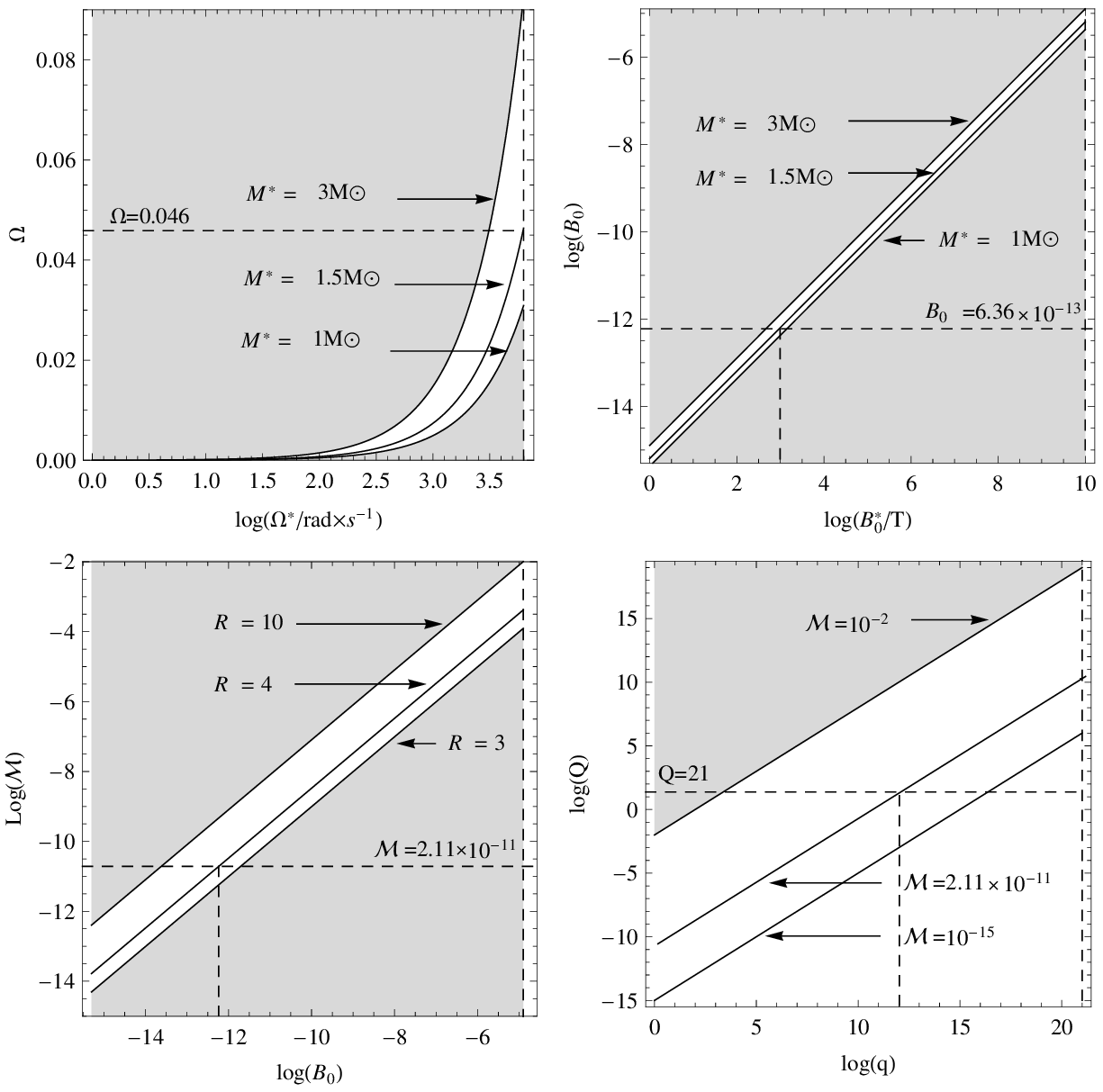}
\caption{Relations between the physical parameters
(magnetic field strength $B_0^*$ and angular velocity of rotation
$\Omega^*$) and the corresponding scaled (dimensionless) parameters,
characterizing a magnetic star and the charged particles moving around
it. The dashed lines denote the assumed limits and some representative values. The gray regions are
astrophysically irrelevant, within the assumed limitation.}
\label{Fig:6}
\end{figure}    
In the upper left plot, starting from the assumed astrophysically relevant values of angular velocity $\Omega^*\leq 10^{3.8}$, we can determine values of $\Omega$ for a fixed mass $M^*$. For example, light magnetic compact stars with $M^*=1\,\rm{M_{\odot}}$ can rotate with the scaled angular velocity up to $\Omega\doteq 0.031$, while the heavy ones with $M^*=3\,\rm{M_{\odot}}$ up to $\Omega\doteq0.093$.  
In the upper right plot, within the astrophysically relevant values of magnetic field strength $B_0\leq 10^{10}$, we can determine values of $B_0$ for a fixed mass $M^*$.\footnote{The  value of $\Omega^*$ is in physical (SI) units,
i.e. rad/s.  Similarly, $B_0^*$ is expressed in units of Tesla.} 
We notice that light magnetic compact stars with $M^*=1\,\rm{M_{\odot}}$
can be endowed with magnetic fields of the scaled strength up to $B_0\doteq 10^{-5.4}$,
while the heavy ones with $M^*=3\,\rm{M_{\odot}}$ up to $B_0\doteq10^{-4.9}$.
In the lower left plot, for the scaled magnetic field strength $B_{0}\leq 10^{-4.9}$, we can determine values of $\mathcal{M}$ for a fixed scaled radius $R$. We can also see that 
small magnetic compact stars with $R=3$ allow scaled magnetic
dipole momenta up to $\mathcal{M}\doteq 10^{-3.9}$, while the large ones with
$R=10$ up to $\mathcal{M}\doteq10^{-2.0}$. 
In the lower right plot, for the specific charges $q\leq 10^{21}$, we can determine values of  $Q$ for a fixed magnetic dipole moment $\mathcal{M}$. Note that small scaled magnetic dipole momenta of $\mathcal{M}=10^{-15}$ allow effective specific charges up to
$Q=10^{6}$, while the big ones of $\mathcal{M}=10^{-2}$ can support values up to $Q=10^{19}$.

For instance, a neutron star with the mass $M^*=1.5\,\rm{M_{\odot}}$
(corresponding to the length-scale  $2212\,{\rm m}$), radius
$R^*=4M^*\doteq 8850\,{\rm m}$ and magnetic field strength
$B_0^*=10^3\,{\rm T}\doteq 2.87\times10^{-16}\,{\rm m}^{-1}$
($\mathcal{M}^*\doteq 1.03\times10^{-4}\,{\rm m}^2$), rotating with the
angular velocity $\Omega^*=6283\,{\rm rad/s}\doteq
2.1\times10^{-5}\,{\rm m}^{-1}$ ($f^*\doteq 10^3\,{\rm Hz}$),
is described by the scaled parameters $\Omega\doteq 0.046$ and
$B_0\doteq 6.36\times10^{-13}$ ($\mathcal{M}\doteq 2.11\times10^{-11}$).
For particles with the specific charge $q=10^{12}$ (charged dust
grains), we have the effective specific charge $Q\doteq 21$ (see figure \ref{Fig:6}).

\subsection{Kerr black hole in uniform magnetic field}
While masses of astrophysical black holes are known with relatively
good precision and customarily categorized in three groups of
stellar-mass black holes ($M^*\lesssim30\,\rm{M_{\odot}}$), intermediate mass
black holes ($10^2\,\rm{M_{\odot}}\lesssim$$M^*$$\lesssim 10^5\,M_{\odot}$), and supermassive black
holes ($M^*\gtrsim10^6\,\rm{M_{\odot}}$), much less is known about the
intensity of cosmic magnetic fields surrounding the black holes.
We assume masses of the black holes in the interval \mbox{$3\,\rm{M_{\odot}}\leq$$M^*$$\leq 10^{10}\,\rm{M_{\odot}}$}~\cite{Zio:2008,Cze-Nik:2009} and the galactic magnetic field strength up to $B_0^*=10^{-6}\,{\rm T}$. Near supermassive black holes with jets the magnetic field is thought to be still significantly stronger~\cite{Rey-Gar-Beg:2006}.

The limits of the used scaled parameters can be determined from the figure \ref{Fig:7}, which can be read in the following way.    
\begin{figure}[t]
\centering
\includegraphics[scale=1.2, trim = 0mm 0mm 0mm 0mm, clip]{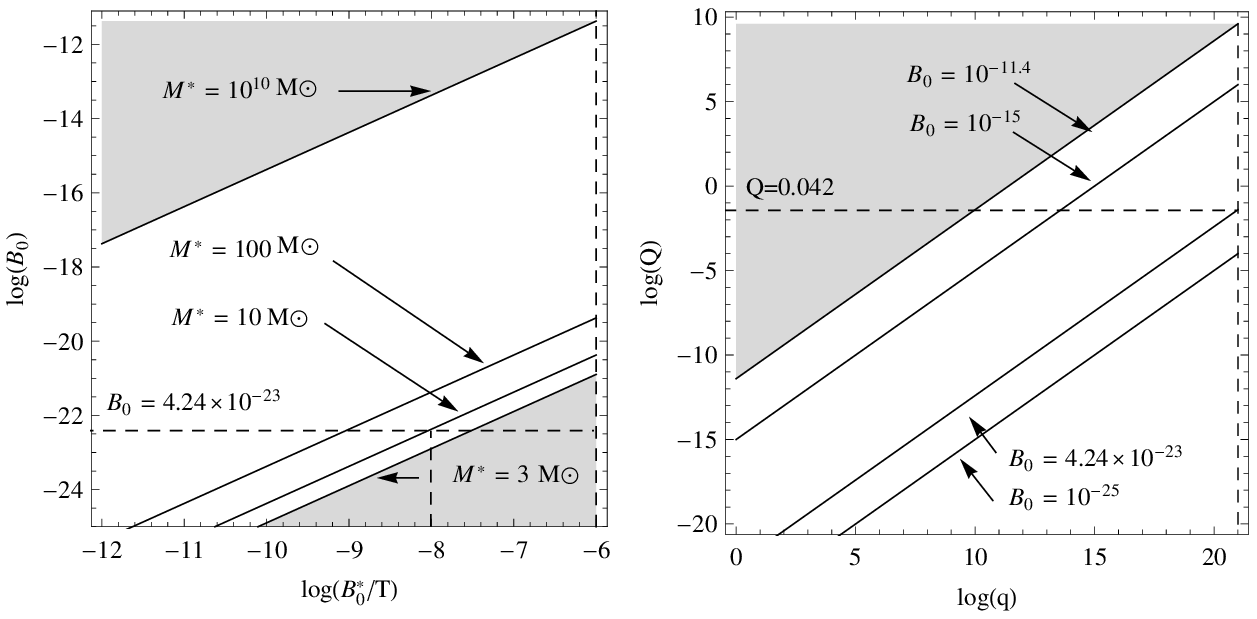}
\caption{As in the previous figure, but for rotating black holes in
the uniform magnetic fields.}
\label{Fig:7}
\end{figure}
In the left plot, within the assumed astrophysically relevant values of the magnetic field strength  $B_0^*\leq10^{-6}$, we can determine values of $B_0$ for a fixed mass $M^*$. We can see that light  black holes with
$M^*=3\,\rm{M_{\odot}}$ allow values of the scaled magnetic field strength up to $B_0\doteq 10^{-20.9}$, while the heavy ones with $M^*=10^{10}\,\rm{M_{\odot}}$ up to $B_0\doteq 10^{-11.4}$. 
In the lower right plot, within the astrophysically relevant values of the specific charge $q\leq 10^{21}$, we can determine values of $Q$ for a fixed scaled magnetic dipole strength $B_0$. We can see that weak magnetic field strengths of $B_0=10^{-25}$ allow values of the effective charge up to $Q=10^{-4}$, while the strong ones of $
B_0=10^{-11.4}$ up to $Q=10^{9.6}$.

For instance, a rotating black hole with the mass
$M^*=10\,\rm{M_{\odot}}\doteq14750\,\rm{m}$ in the asymptotically uniform
magnetic field of the strength $B_0^*=10^{-8}\,{\rm
T}\doteq2.87\times10^{-27}{\rm m}^{-1}$ is described by the parameter
$B_0\doteq 4.24\times10^{-23}$. For particles with the specific charge of magnitude
$q=10^{21}$ (electrons), we have the scaled effective charge
$Q\doteq0.042$ (see figure \ref{Fig:7}). Realistic values for the scaled spin $a$ are
limited by the value $a=1$. For the greater values of the spin, the
black-hole horizons disappear and the Kerr metric describes the
naked-singularity spacetime. 

\subsection{Possible applications}
Two kinds of charged particles seem to be relevant with respect to the
halo orbits. Firstly, in the case of electrons and protons, the magnitude of specific
charges has a large value, and so the halo orbits can exist only very
close to the black hole horizon or near an ultra-compact magnetic star.
This kind of motion has been recently explored in the context of
magnetospherical motion in stellar-mass black hole binaries~\cite{Takahashi09}. Magnetically driven oscillations of matter near
black holes have recently been explored also in the context of equatorial
motion~\cite{Bak-etal:2010}. 

Another promising application of the halo motion concerns charged dust
grains which typically acquire the magnitude of specific charge much smaller than that of 
elementary particles. Therefore, the gravitational effects
influence the motion of dust grains more efficiently, and so they can
form the halo orbits at larger distances from the star surface. Infrared
observations are suitable to test the possible existence of halo orbits
of dust grains around  magnetic compact stars~\cite{Wang06}. The debris
discs of such particles  could perturb the electromagnetic signal of the
compact star and give rise to  quasi-periodic modulation.

It is also well known that the frequencies of epicyclic geodesic motion~\cite{Ali-Gal:1981,Tor-Stu:2005} have a substantial role in the resonant (or other) models of high-frequency quasiperiodic oscillation observed in the black hole~\cite{Tor-etal:2005,Stu-etal:2005,Stu-Sla-Tor:2007} and neutron star~\cite{Tor-etal:2008} low mass X-ray binaries, and their magnetic-field induced modification could be of high importance, as shown in the case of equatorial motion~\cite{Bak-etal:2010}.

\section{\label{sec:Conclusions}Conclusions}
Our calculations reveal the properties of halo motion of charged
particles near magnetic compact stars and black holes. These can be
relevant e.g. for charged dust grains when they acquire a small magnitude of  
electric charge and occur in the strong gravitational field. Such
particles can originate from supernova material which falls back and
migrates into the pulsar vicinity before evaporating.

The halo orbits can take place in axially symmetric systems: near
a magnetic compact star, which we modelled by the Schwarzschild geometry
and a test rotating dipole magnetic field, as well as near a Kerr black
hole immersed in an asymptotically uniform
magnetic field. 

In this paper, we concentrated ourselves on the methodological aspects of
the halo orbits and their systematic classification. Our main interest
concerned the properties of the halo orbits in situations when the
motion takes place in a strong gravitational field. We only briefly
touched the astrophysical applications and will investigate them further
in the following work. Table \ref{Tab:3} summarizes the basic
characteristics of different  examples of strongly gravitating systems
which exhibit the halo orbits.
\begin{table}[tbh!]
\caption{Orientation of motion of electrically charged particles 
along stable halo orbits and the corresponding sign of the charge. The
results correspond to the configuration where the magnetic field
(characterized by the magnetic  dipole moment $\mathcal{M}$ or magnetic
field strength in the equatorial  plane $B_0$) is oriented in the same
direction as the rotation  vector, characterized by the signs of
$\Omega$ or $a$.}   {\small\vspace*{4mm}
\begin{indented}
\item\begin{tabular}{@{}lll}
\br\rule[-1em]{0pt}{2.5em}
Compact object&Charge of particles&Orientation of motion\\
\hline
\hline\rule[0em]{0pt}{1.5em}
Rotating magnetic star & Positive & Co-rotation\\
                     & Positive & Counter-rotation\\   
\rule[-1em]{0pt}{1.5em}                     & Negative  & Co-rotation\\
\hline\rule[0em]{0pt}{1.5em}
Static magnetic star & Positive & Counter-rotation\\
\rule[-1em]{0pt}{1.5em}   & Negative  & Co-rotation\\
\hline\rule[0em]{0pt}{1.5em}
Magnetized black hole & Positive & Counter-rotation\\
\rule[-1em]{0pt}{1.5em} (Kerr metric) & Negative  & No-orbits\\
\br
\end{tabular}
\end{indented}
\label{Tab:3}}
\end{table}

As for the magnetic compact stars, results of our
discussion of the stable halo orbits orientation are qualitatively
consistent with those obtained previously in Newtonian and the
pseudo-Newtonian investigations~\cite{Dull-Hor-How:2002,Kov-Stu-Kar:2008}.
We notice that the strong gravitational field and the presence of
event horizon affect the form of the effective potential and introduce
significant changes  with respect to the Newtonian case. However, we
also notice that the  pseudo-Newtonian approach can give surprisingly
good information on the loci of halo orbits in spherically symmetric
gravitational fields of magnetic stars, in which case the results
qualitatively agree with the  exact general relativistic approach
presented here. On the other hand, the pseudo-Newtonian modelling of
Kerr spacetimes~\cite{SemK99,Muk:2002}  is much more complicated, and
seems to be less fruitful for modelling the phenomena in the Kerr
geometry, which itself is relatively simple. 

The effective potential of the halo motion exhibits a number of
qualitatively different types. Two basic categories can be
distinguished. The first one allows particles, bounded within the halo
potential lobes, to commute across the equatorial plane. On the other
hand, the second category does not allow such an interconnection, as the
two lobes are entirely disjoint. The off-equatorial halo motion is an
interesting phenomenon not only because it has not yet been explored in
full detail in the literature, but also because it shows a variety of
orbits, depending on orbital parameters: off-equatorial stable circular
motion which does not cross the equatorial plane; small oscillations in
the radial and vertical directions, with characteristic frequencies
around the circular orbits; and the motion that traverses
between the lobes when the energy is increased to sufficiently high
levels. Recently, the off-equatorial motion of charged particles near Kerr black hole in asymptoticaly uniform magnetic field has been discussed from a different point of view in the ref. \cite{Pre:2010} as well.

\ack
The authors thank M. Urbanec for useful discussions concerning section 5.1.
Institute of Physics and the Astronomical Institute have been operated
under the projects  MSM\,4781305903 and AV\,0Z10030501, and further
supported by  the Centre for Theoretical Astrophysics LC06014 in the
Czech Republic. JK, VK and ZS thank the Czech Science Foundation (ref.\ P209/10/P190, 205/07/0052, 202/09/0772). OK acknowledges the doctoral student
program of the Czech Science Foundation (ref.\ 205/09/H033).

\end{document}